\documentclass[11pt,twoside]{article}
\usepackage{newpasp}
\usepackage{epsf}

\index{Mashchenko, S.}
\index{St-Louis, N.}

\begin{document}

\title{Automatic Shell Detection in CGPS Data}
\author{Sergey Mashchenko \& Nicole St-Louis}
\affil{D\'epartement de Physique and Observatoire du Mont M\'egantic,
 Universit\'e de Montr\'eal, C.P. 6128, Succursale Centre Ville,
 Montr\'eal, H3C 3J7, QC, Canada}

\begin{abstract}
A numerical code aiming at the automatic detection of spherical expanding HI shells
in radio
 data-cubes is presented. The following five shell parameters
are allowed to vary within a specified range: angular radius, expansion velocity
and three center coordinates (galactic coordinates $l$ and $b$, and systemic 
radial velocity $V$).
We discuss several factors which can reduce
the sensitivity of the shell detection: the presence of noise (both instrumental
and ``structural''), the fragmentation and asphericity of the shell and the 
inhomogeneity of 
the background and/or foreground emission. The code is tested on four objects:
two HII regions and two early B stars. 
We use HI data from 
the Canadian Galactic Plane Survey.
In all four 
cases the evidence for an expanding HI shell is found.
\end{abstract}

\section{Introduction}

Shells are omnipresent in the interstellar medium (ISM) of spiral and dwarf irregular
galaxies. In some systems most of the ISM can be described as an ensemble of interacting
shells and supershells (Staveley-Smith et al. 1997; Mashchenko, Thilker, \& Braun 1999,
hereafter MTB).

Theoretically, many astrophysical phenomena  may lead to the formation of
expanding HI shells:
stellar winds (Weaver et al. 1977), supernova explosions (Cox 1972), the combined action
of multiple supernovae and stellar winds in OB associations (Bruhweiler et al. 1980),
radiation pressure from the field stars (Elmegreen \& Chiang 1982)
and the infall of high-velocity clouds onto the
galactic plane (Tenorio-Tagle 1981).  In more general terms, any physical mechanism
which is able to drive an expanding supersonic shock wave into the surrounding ISM long enough
for radiative cooling in the shell of swept-up gas to become significant, will lead
to the formation of a relatively thin ($\la$10\% of the shell radius) and
cold ($\sim$ 100--1000~K) expanding HI shell. (See Bisnovatyi-Kogan \& Silich 1995 for a review of 
shock wave propagation in the ISM.) 

Unfortunately, there are many factors which can complicate the above picture and
make the HI shells more difficult to detect. The blowing out of an interstellar bubble
caused by the presence of a sharp density gradient in the surrounding medium makes the shell
susceptible to Rayleigh-Taylor instabilities, which results in its fragmentation.
The propagation of the shock through the clumpy turbulent ISM
can be another destructive factor.
The shells formed around O and early B stars are expected to be at least
partially ionized by the ionizing photons from the star (Weaver et al. 1977). 
The stratification of the ISM structure
and the differential rotation of the galactic disk result in the highly aspherical appearance
of evolved supershells (Mashchenko \& Silich 1994; Silich et al. 1996).
The detection of expanding HI shells
in edge-on disk systems and in our own Galaxy is further hampered by the presence of confusing
emission along the line of sight.

As a result, few HI shells have been detected around stars with strong stellar wind.
There are $\sim$30 HI shell detections for WR stars 
(Arnal \& Mirabel 1991, Dubner et al. 1992, Pineault et al. 1996,
Arnal et al. 1999, Cappa et al. 1999, Gervais \& St-Louis 1999, and references therein),
and just a few shell candidates for O stars 
(van der Bij \& Arnal 1986; Benaglia \& Cappa 1999 and references therein) and
B stars (Dubner et al. 1992).

Pearson correlation coefficient based algorithms 
can be successfully used
for the automatic recognition of complex kinematic structures in the radio data-cubes
(Thilker, Braun, \& Walterbos 1998; MTB). A cross-correlation approach is tolerant
to the partial fragmentation and ionization of the shell, to the mild distortion of its
shape and to the presence of slowly varying background emission (MTB).

In this article we describe an object recognition code for the automatic detection
of the expanding HI shells. The code uses the Pearson correlation coefficient
as a measure of similarity between the data and a simple model of the shell.

\section{Model}

We assume that all the swept-up gas is located in a relatively thin spherical shell.
If $R$ is the radius of the shell and $\rho_0$ 
is the gas density in the surrounding medium, then the total
shell mass is equal to $M=4/3 \pi R^3 \rho_0$. Alternatively, 
$M \simeq 4 \pi R^2 d \rho_{sh}$ for $\delta \equiv d/R \ll 1$, 
where $\rho_{sh}$ is the density
of the shell gas and $d$ is the thickness of the shell. Then 
$\delta \simeq \rho_0/(3 \rho_{sh})$. The gas density jump condition
for an adiabatic shock gives us an estimate for the density ratio:
$\rho_0/\rho_{sh} \simeq [(\gamma -1)M^2+2]/[(\gamma +1)M^2]$, where
$\gamma$ is the adiabatic index and $M$ is the Mach number for the shock
(Landau \& Lifshitz 1959, p.~331). Finally, for a monoatomic gas ($\gamma =5/3$)
we obtain $\delta \simeq 1/12$ for a strong shock ($M\to \infty$)
and $\delta \simeq 1/7$ for a relatively weak $M=2$ shock.

The relative shell thickness values $\delta$ thus obtained should be considered
as an upper limit for the shells significantly affected by the radiative cooling.
Therefore we adopt a constant value of $\delta=1/10$ in our code as a reasonable
guess for both mildly supersonic radiative shocks and adiabatic shocks.

\section{Automatic recognition algorithm}

We consider the simple situation for which the radial velocity of an object
(HII region, O or B star,
etc.) is known. In this case the direct computation of the correlation coefficient for
all possible values of the parameters is feasible. We vary explicitly the 
following five shell parameters:
radius $R$, expansion velocity $V_e$ and image pixel 
coordinates of the shell center $X_0$, $Y_0$ and $Z_0$ corresponding to the
galactic coordinates $l_0$ and $b_0$, and systemic radial velocity of the shell $V_0$.
To restrict the range of $X_0$, $Y_0$ and $Z_0$, we demand that
our object be located within the shell boundaries: 
$[(X_0-X_*)^2+(Y_0-Y_*)^2]/R^2 + (Z_0-Z_*)^2/V_e^2 \le 1$, where $X_*$, $Y_*$ and $Z_*$
are the pixel coordinates of the object, and $R$ and $V_e$ are in image pixel units.

The linear invariance of the Pearson correlation coefficient $\rho$ introduces implicitly
two additional free parameters: a scale (corresponding to the integral brightness
of the shell) and an additive constant (corresponding to the level of the background
and/or foreground emission $B$). To deal with more complex background distributions we
preprocess the HI data-cube by removing a linear dependency of $B$ on the spatial 
coordinates $X$ and $Y$  within the search area --- 
independently for each velocity channel.

The sensitivity of the shell detection is limited by the presence of noise in the data.
Tests show that for all but the smallest shells, the ``structural'' noise
(caused by the filamentary structure of the ISM) dominates over instrumental noise
and is a function of the number of non-zero pixels
in the projected model cube $N$. We apply the code to a few ``quiescent'' fields in 
the Canadian Galactic Plane Survey (CGPS, e.g. English et al. 1998) 
HI data which contain no known shell generating objects
(O, B, WR stars, SNRs, HII regions). The results of the test are used to draw a 99\% 
confidence curve $\rho_{99}(\log N)$. The normalized correlation coefficient
$\rho/\rho_{99}$ has a constant signal-to-noise ratio for any $\log N$ value.
High probability shell detections have $\rho/\rho_{99}>1$.

The shell detection procedure consists of three basic steps: (1) The HI data-cube
is preprocessed to remove the linear dependency of $B$. (2) 
For each combination of $R$ and $V$
the projection of the shell model into the observational frame of reference 
$X$--$Y$--$Z$ is performed. The model is then cross-correlated with the data for
all possible 3D translations and the highest value obtained of the correlation
coefficient $\rho$ is stored along with the corresponding values of $R$, $V$, $N$
and translation vector $\{\Delta X,\Delta Y,\Delta Z\}$.
(3) The quantity $\rho/\rho_{99}$ is plotted against $\log N$. The biggest 
$\rho/\rho_{99}$ value corresponds to the most probable shell detection. 

Tests confirm the ability of the code to find an expanding spherical shell
in a realistic environment. When we add model shells with different integral brightness to
real HI data-cubes and apply the code, we find that the shell can be 
reliably detected
even if it is too weak to be seen by visual inspection of the cube.

\section{Application of the code to CGPS data}

\begin{figure}[p]
\plotfiddle{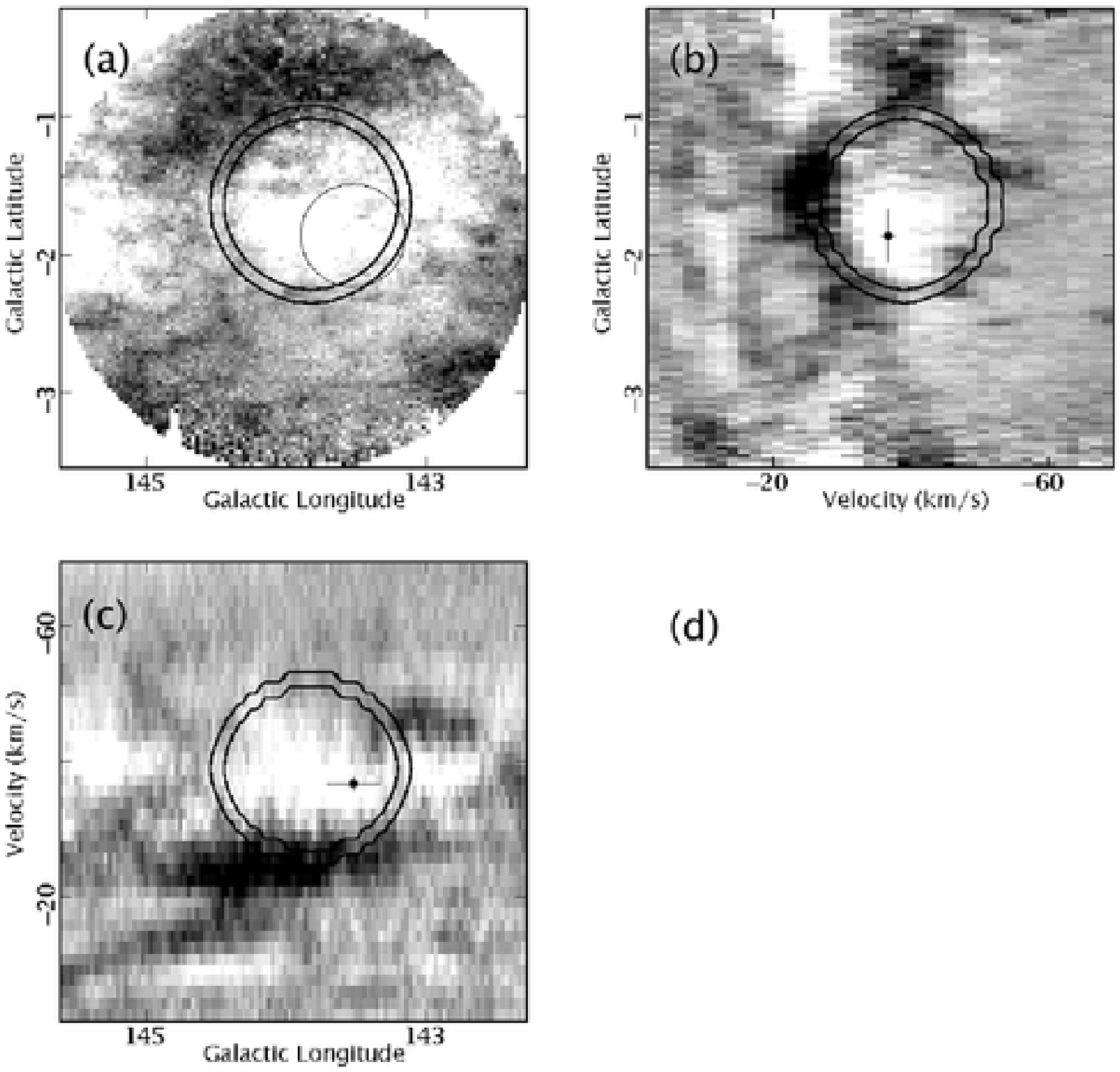}{203pt}{0}{51.1}{51.1}{-160.125}{-100.5}
\plotfiddle{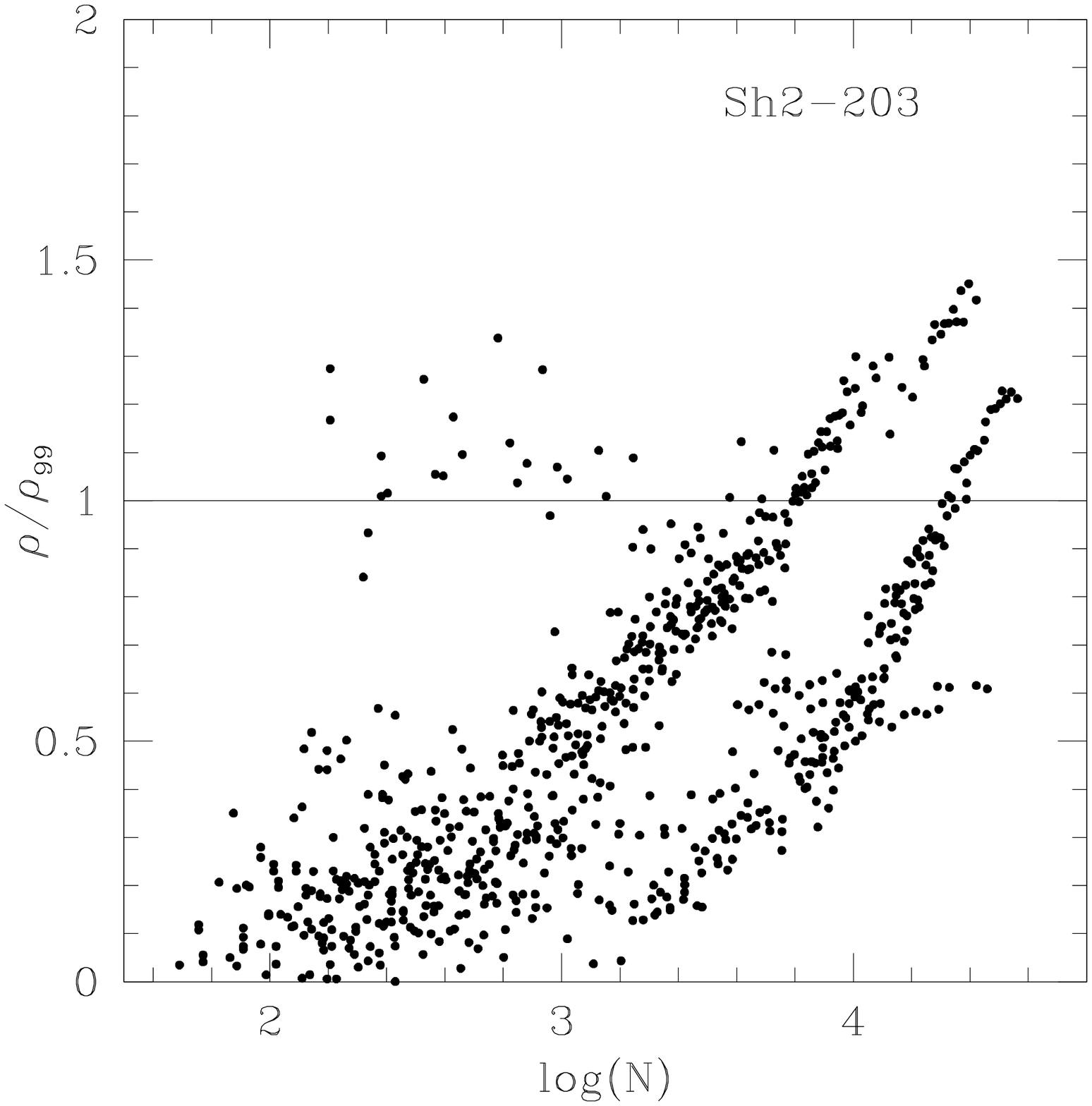}{0pt}{0}{20.5}{20.5}{5}{-25}
\caption{Shell candidate for Sh2--203. 
The HII region is marked as a thin line circle (panel (a)), 
and as a cross (panels (b) and (c)). The thick line ring
corresponds to the best fitting model (Table~1).}
\end{figure}
\begin{figure}[p]
\plotfiddle{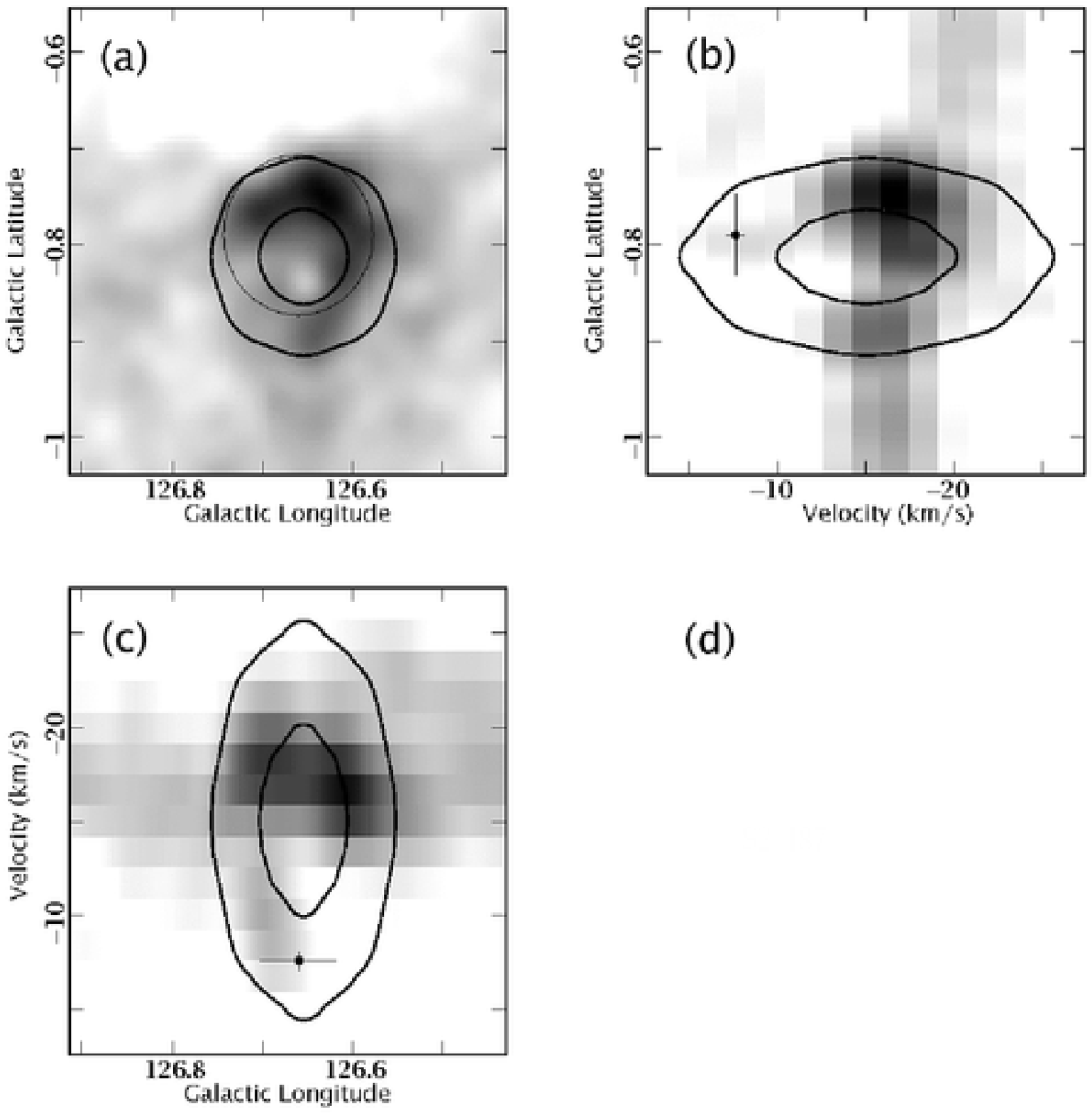}{215pt}{0}{50}{50}{-158.5}{-92}
\plotfiddle{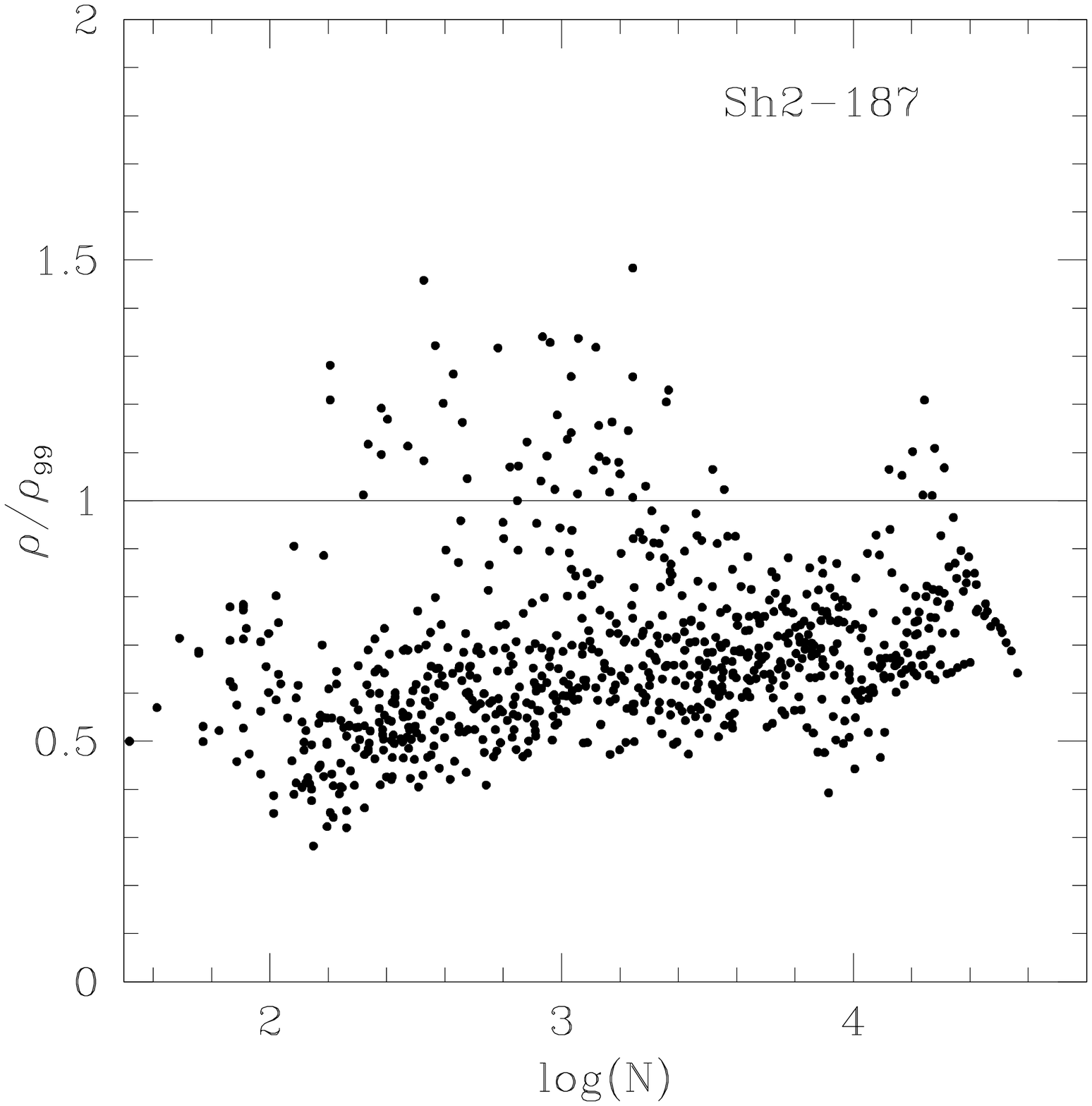}{0pt}{0}{20.5}{20.5}{5}{-24}
\caption{Same as Fig.~1, but for the case of Sh2--187.}
\end{figure}
\begin{figure}[p]
\plotfiddle{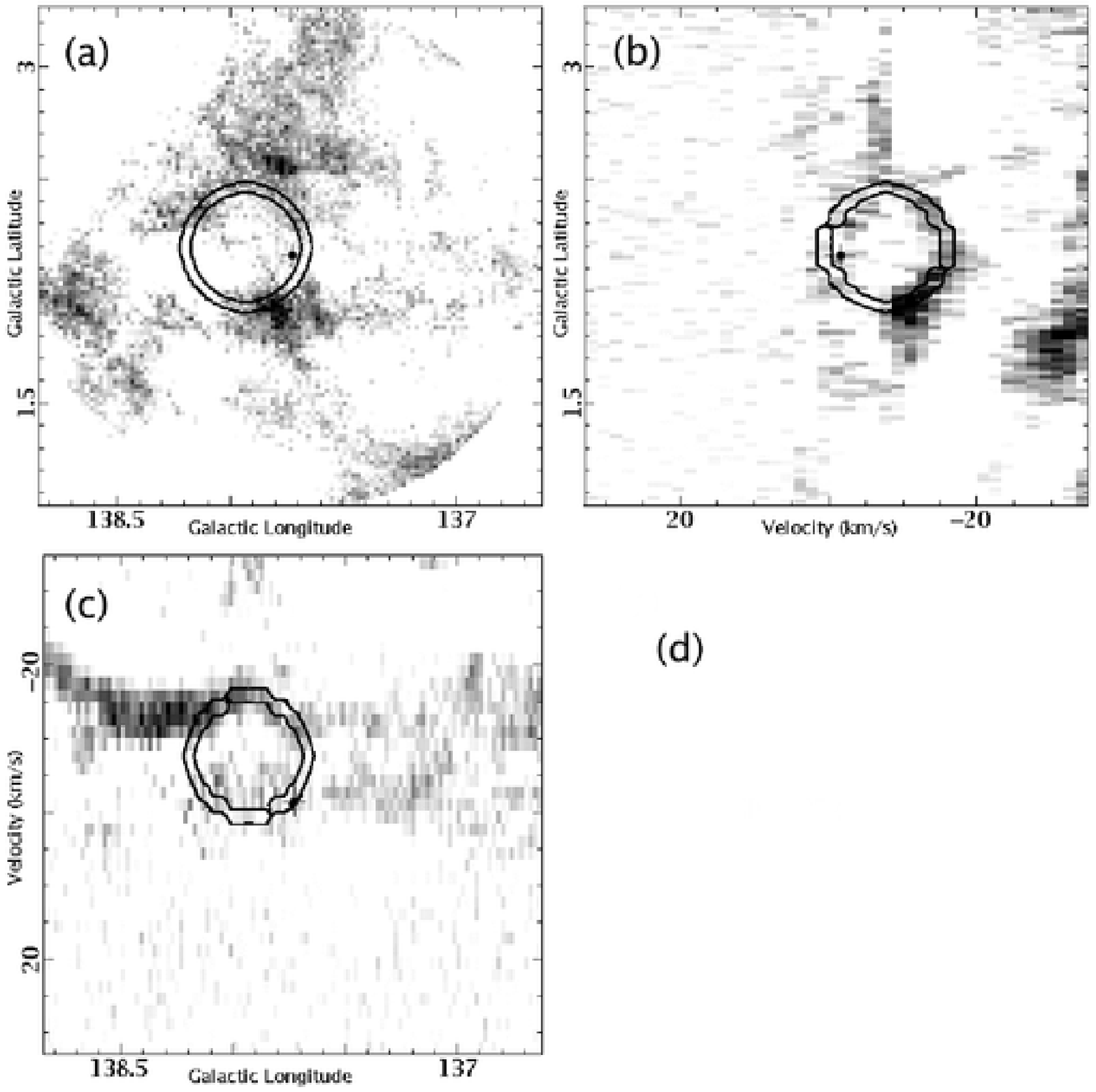}{215pt}{0}{50}{50}{-157.5}{-92.5}
\plotfiddle{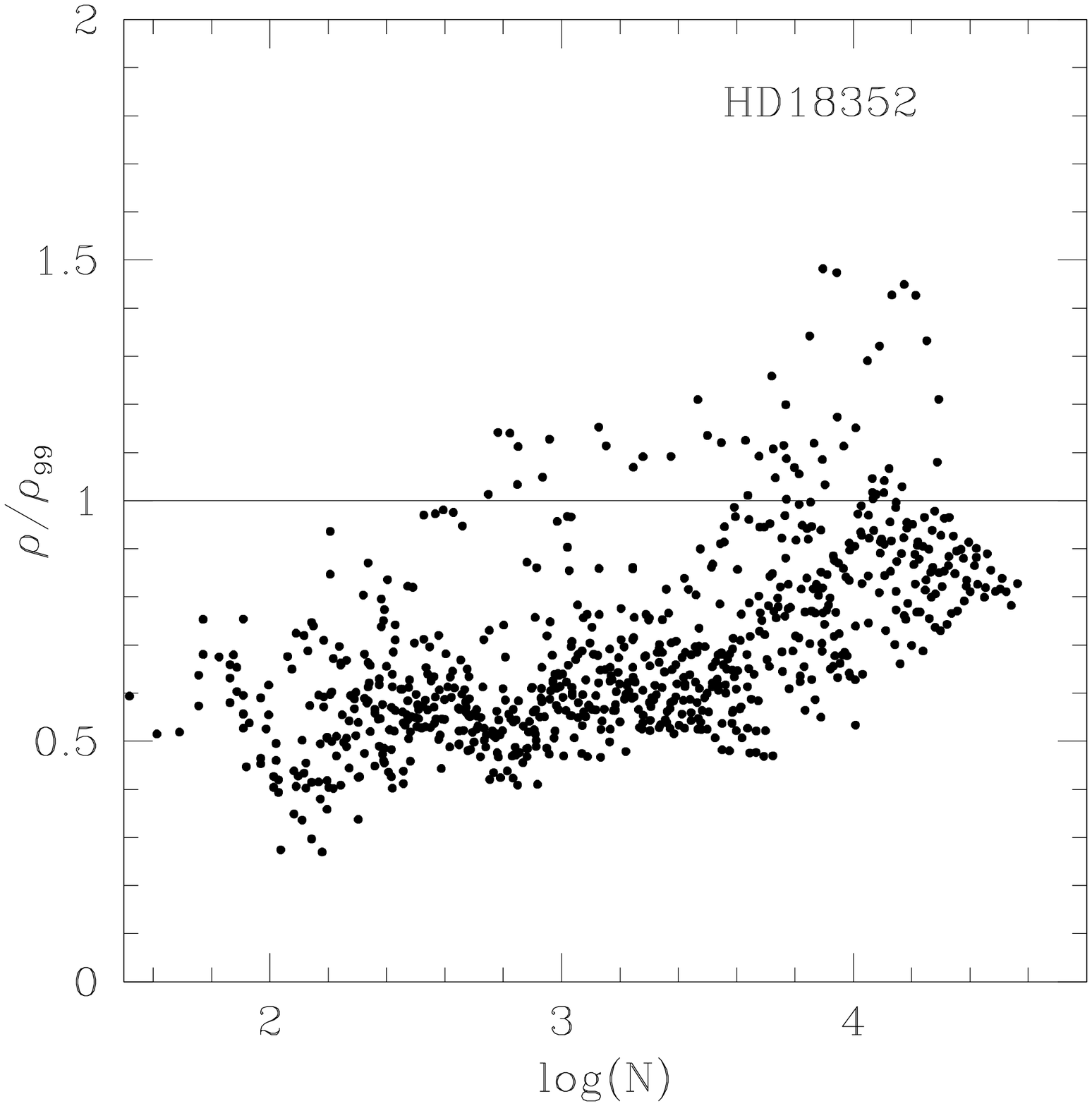}{0pt}{0}{22}{22}{-2}{-35}
\caption{Shell candidate for HD18352. 
The star is marked as a cross. The thick line ring
corresponds to the best fitting model (Table~1).}
\end{figure}
\begin{figure}[p]
\plotfiddle{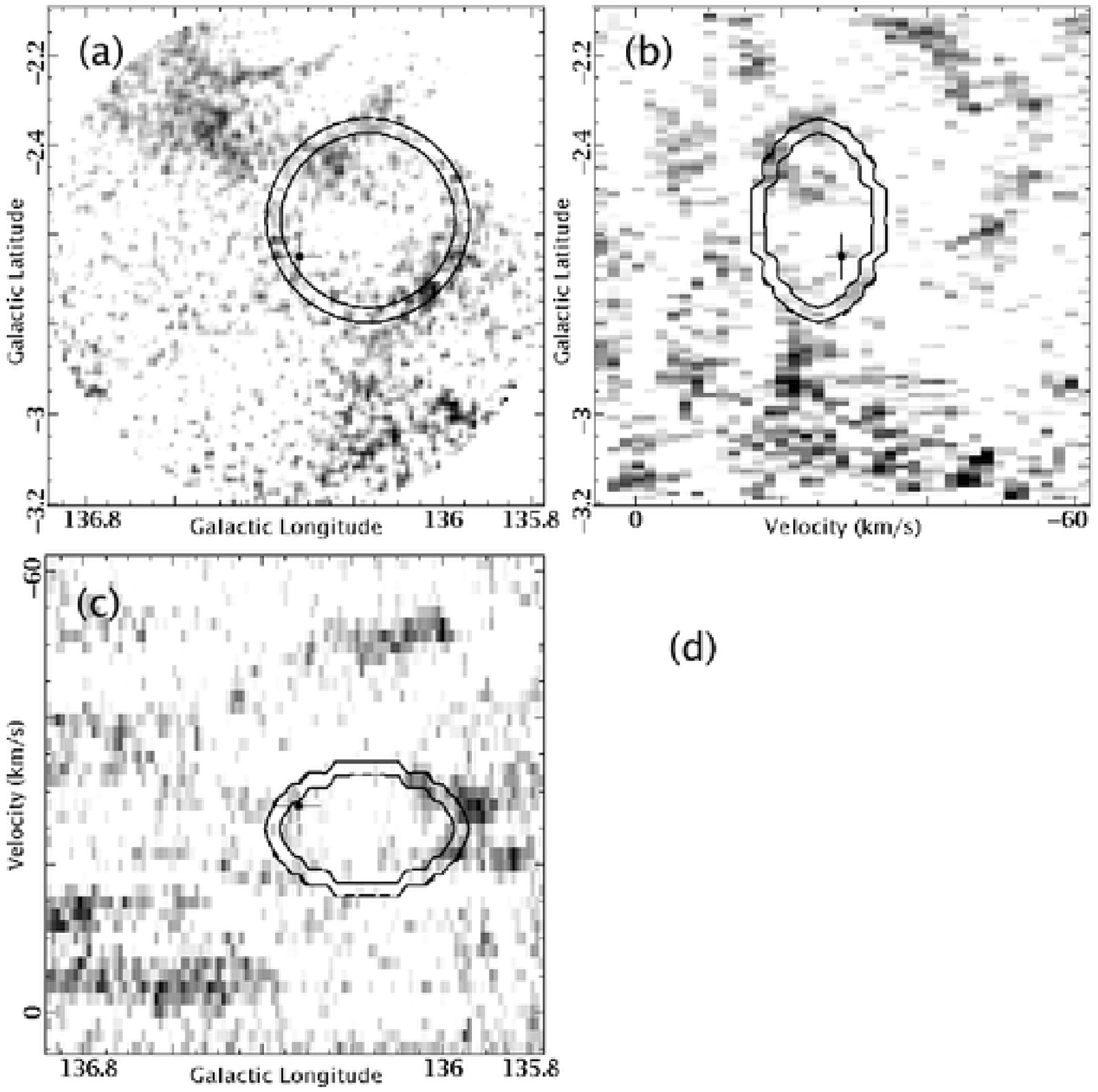}{215pt}{0}{50}{50}{-157.5}{-92.5}
\plotfiddle{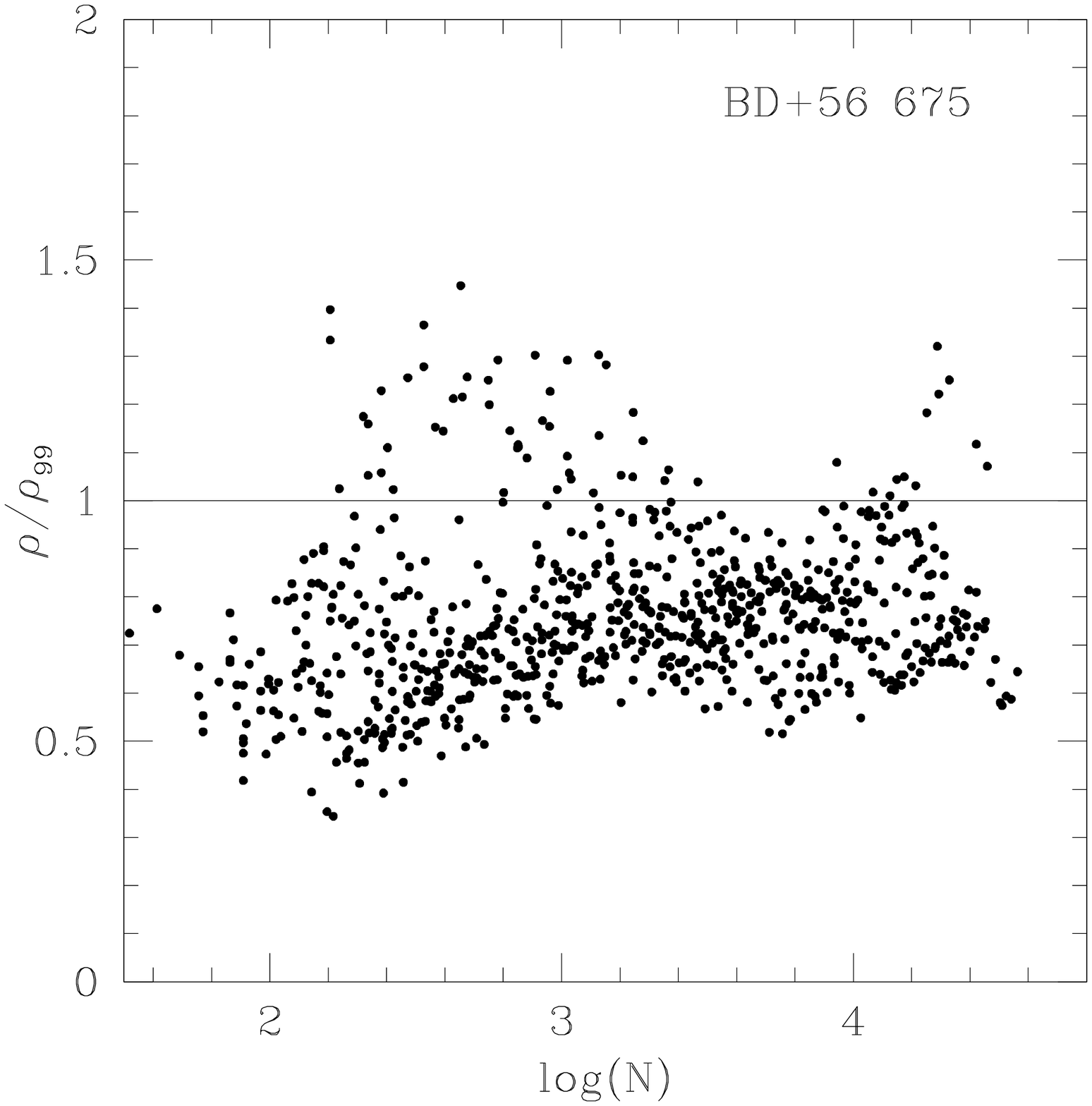}{0pt}{0}{22}{22}{-2}{-35}
\caption{Same as Fig.~3, but for the case of BD+56~675.}
\end{figure}

We have tested the shell detection code on four objects: two HII regions
(Sh2--187 and Sh2--203) and two B1V stars (HD18352 and BD+56~675).
The HII regions cover a wide range of sizes ($10'$ for Sh2--187,
and $45'$ for Sh2--203 --- Sharpless 1959).  Sh2--187 is known to possess
an HI shell (Joncas, Durand, \& Roger 1992). Radial velocity values for all four
objects are available (Fich et al. 1990; SIMBAD).
We have used the CGPS HI data,
which have high spatial ($\sim 1'$) and velocity (1.32~km~s$^{-1}$) resolution.
The search was limited to shells with  an expansion velocity $V_e$ in the 
range 3.3~km~s$^{-1}\le V_e \le 15$~km~s$^{-1}$ and an angular radius $R$ in the range
$1.'2 \le R \le 16'$ (HD18352 and BD+56~675), 
$2' \le R \le 26'$ (Sh2--187) and $3.'6 \le R \le 47'$ (Sh2--203).

In all four cases shell candidates with $\rho/\rho_{99}>1.3$ were found.
Figures~1--4 show the detection plots (panels (d)) and three 
orthogonal slices through 
the preprocessed data-cubes (panels (a)--(c)). In most cases there is evidence for
a multi-shell structure (Figures 1d, 2d, 4d). The data on
the highest probability
shell detections are summarized in Table~1. 

\begin{center}
{\small
\begin{table}[t]
\caption{Parameters of the shell candidates}
\vspace{3pt}
\begin{tabular}{lcccccccc}
\tableline
\tableline
Object    & log(N) & $\rho/\rho_{99}$ & $l_0$, & $b_0$,  & $V_0$,    & $R$,   & $V_e$,        \\
          &        &                  & dgr.   & dgr.    &km~s$^{-1}$& arcmin & km~s$^{-1}$  \\
\tableline
Sh2--203  & 4.40   & 1.45             & 143.82 & $-1.63$ & $-39$     & 42     & 14          \\
Sh2--187  & 3.24   & 1.48             & 126.65 & $-0.81$ & $-15$     & 5.4    & 9            \\
HD18352   & 3.90   & 1.48             & 137.93 & $+2.20$ & $-8$      & 17     & 9           \\
BD+56 675 & 4.29   & 1.32             & 136.17 & $-2.57$ & $-25$     & 13     & 9            \\
\tableline
\tableline
\end{tabular}
\end{table}
}
\end{center}

\vspace{-10mm}

\acknowledgments
We would like to thank Francis Boulva for the help with the data processing. 
This work is supported by a grant from the Natural Sciences and
Engineering Research Council of Canada. 


\end{document}